%% file: capuzzodolc.tex
\documentclass[multphys,vecphys]{svmult}

\usepackage{makeidx}         
\usepackage{graphicx}        
\usepackage{multicol}        
\usepackage[bottom]{footmisc}

\makeindex             

\def\beq{\begin{equation}}
\def\eeq{\end{equation}}

\def\undersim#1{\mathop{\vtop{\ialign{##\crcr
$\hfil\displaystyle{#1}\hfil$\crcr\noalign
{\kern0pt\nointerlineskip}\hbox{$\hfil\sim\hfil$}\crcr
}}}}

\begin{document}

\title*{Globular Cluster System evolution in early type galaxies}
\author{R. Capuzzo-Dolcetta\inst{1}}
\institute{Dipartimento di Fisica, Universit\'a di Roma ``La Sapienza",
P.le Aldo Moro, 2, I-00185 -- Rome, Italy
\texttt{roberto.capuzzodolcetta@uniroma1.it}}
%
%
\maketitle
\abstract
Globular clusters (GCs) constitute a system which is evolving because of
various interactions with the galactic environment. Evolution may be the
explanation of many observed features of Globular Cluster Systems (GCSs); the
different radial
distribution of the GCS and the stellar component of early type galaxies is
explained by
dynamical friction and tidal effects, this latter acting both on the large
scale (that of the bulge-halo stars) and on the small scale
(that of the nucleus, often containing a central massive black hole). Merging of
quickly orbitally decayed massive GCs leads to formation of a Super Star Cluster
(SSC)
which enriches the galactic nucleus and is a reservoire of mass-energy
for a centrally located black hole.

\section{Introduction}
\label{intro}
The Hubble Space Telescope and large ground based telescopes 
are providing a continuosly increasing amount of data
concerning GCSs in galaxies, mainly of the early types,
since the pioneering work \cite{har79}.

Two are the most debated points: (i) the difference in
the GCS and galaxy light spatial distribution,
and, (ii) the existence of a bimodal color distribution for GCSs,
and the possible differences between the {\it blue} and the 
{\it red} population,
\newline The two points are, likely, related;
in any case, here I will not discuss about point (ii) 
(see the recent \cite{pen06} paper) but just about point 
(i) which is better observationally stated and deserves a 
correct interpretation.


\section{The GCS and stellar radial distributions in galaxies}
\label{GCSrdis}
It is nowadays clear that the majority of galaxies shows a  radial 
profile of their GCS shallower than that of the stars toward
the galactic centre. Ellipticals show a more or less peaked stellar profile
toward the galactic center (actually, many have a \lq cuspy\rq~ profile), 
while the GCS radial distribution has, usually, a core. The 
related literature is so vast that we limit to recall 
\cite{gri94},\cite{for96}. The explanation of this difference 
in terms of formation and evolution of elliptical galaxies 
(see \cite{har86}, \cite{for96}, \cite{ash98}, \cite{bek06}).
or in terms of evolution of the GCS itself (see \cite{cap93},
\cite{captes99}, \cite{cap05}) is still debated.
\newline The interpretation on the basis of GCS evolution is 
more appealing, because much simpler and not based on 
qualitative and arbitrary modelizations of GC formation in 
galaxies (remember the Occam's razor...).
Moreover, it has other important astrophysical implications. 

Why is it simple? Because it is  based just on the, conservative, 
assumption that the GCS and the halo-bulge galactic stars are 
coeval and had initially the same spatial distribution; 
the presently observed difference can be caused by evolution of 
the GCS. 
That GCSs in galaxies undergo to evolution is undoubtful, 
because they are evolving aggregates of stars moving in an 
external potential which influences the system also by tidal 
distortion and by dynamical friction.
A detailed analysis of the GCS radial profile evolution in early type
galaxies has been presented in \cite{captes99} where a convincing 
explanation of the observed comparative features of GCS and stellar 
light profiles is given.
\par
\begin{figure*}
   \centering
   \resizebox{\hsize}{!}
   {\includegraphics[height=16cm]{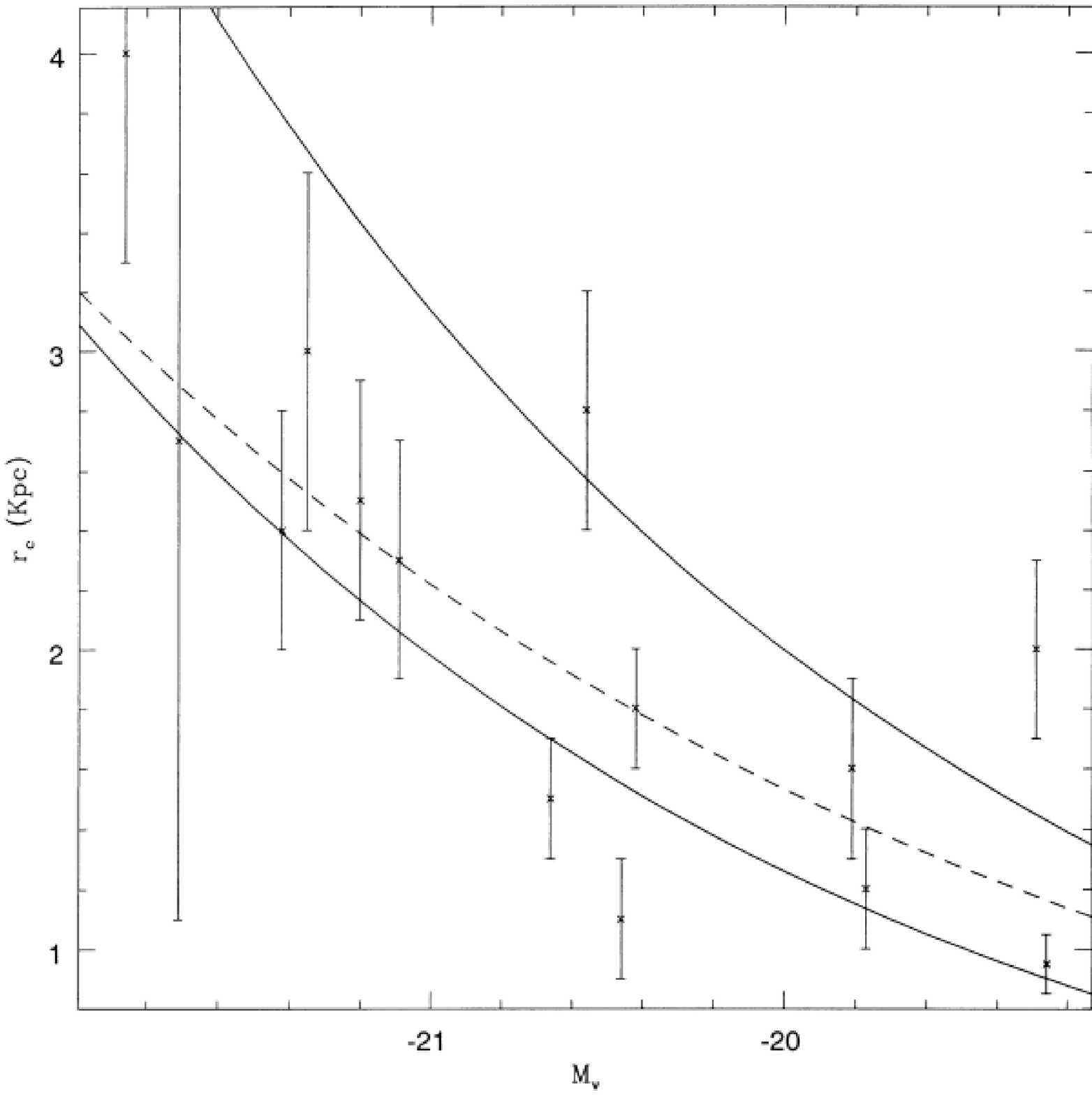}
   \includegraphics[height=16cm]{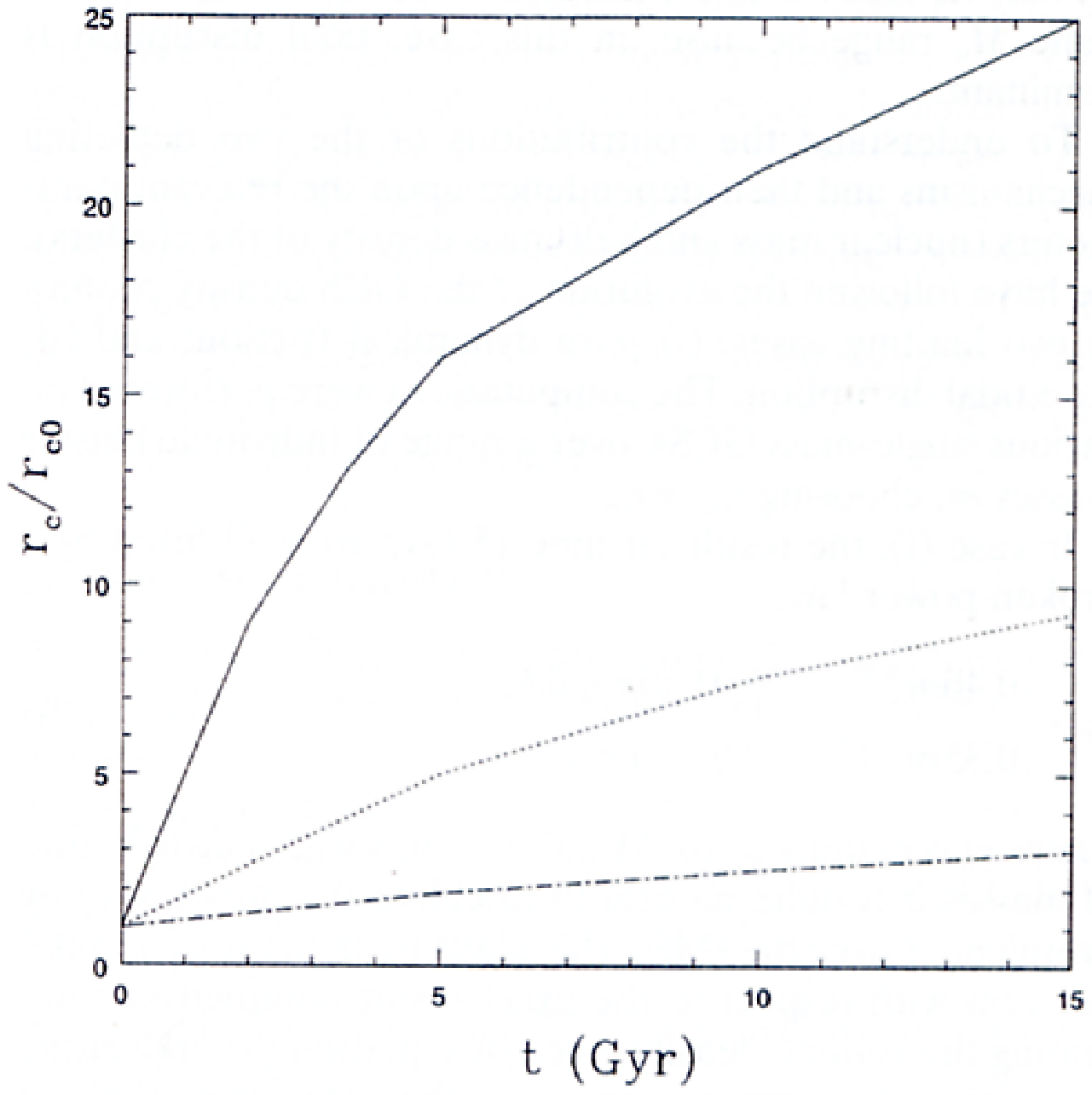}}
     \caption{Left panel: GCS core radius as a function of the absolute 
integrated V mag. of the parent galaxy. Dots
     refer to data in \cite{for96}, with their best fit as dashed line. Solid 
lines are two evoltionary models, with
     two different initial value of the GCS core radius (see Fig. 3 in 
\cite{captes99}).
     Right panel: time evolution of the core radius of the GCS in a triaxial 
galaxy containing a central 
     BH of mass (from bottom up) $10^7$, $10^8$, $10^9$ M$_\odot$ (Fig. 6 in 
\cite{captes97}).}
        \label{scales}
    \end{figure*}

 Some researchers have invoked one observational 
feature, the GCS radial distribution  being  shallower for brighter
galaxies than for faint \cite{for96},
as evidence against the \lq evolutionary\rq ~ explanation.
\newline Apart from that the claimed correlation is not universal 
(for instance, \cite{bea06} found a quite shallow GCS radial 
distribution in the Virgo dE VCC 1087), the evolution 
of a GCS due to the combined role of
dynamical friction, acting on the large scale, and nuclear 
tidal distortion, on a smaller scale, leads to a 
correlation between the slope of the GCS radial profile
and the galaxy integrated luminosity exactly as observed (see Fig. 
~\ref{scales}, left panel). 
This because of the existing correlation
between the two scales through the positive galaxy mass-central black hole (BH) 
mass in galaxies. As example of observational output of the large (GCS core 
radius) and small
(BH tidal destruction radius) scale-correlation see 
right panel of Fig. ~\ref{scales}. 
In conclusion, the GCS slope vs. galaxy luminosity correlation is not, 
unfortunately, a way to distinguish between the two above mentioned hypotheses 
(compare left panel of Fig. ~\ref{scales} with Fig. 4 in \cite{bek06}).
\section{Super star cluster formation and nucleus accretion}
\label{ssc}
There is growing evidence of the presence of very massive young clusters,
as extensively discussed in this Conference, up to the extremely large mass of 
W3 in 
NGC 7252 ($M = 8 \pm 2 \times 10^7$ M$_\odot$, \cite{mar04}).
Massive clusters are not an insignificant fraction of the GCs in galaxies; on 
the contrary,
\cite{har06} indicates how up to a 40\% of the total mass in the GCS of 
brightest cluster galaxies 
is contributed by massive GCs (p.d. mass $> 1.5\times 10^6$ M$_\odot$), in good  
agreement with
recent theoretical results by \cite{kra05}.
\par The initial presence of massive clusters in a galaxy makes particularly
intriguing the GCS evolutionary frame sketched in Sect.2, for the presence of 
some
massive primordial custers may have had very important consequences on the 
initial evolution of the
parent galaxy.
Actually, the GCS evolution in an elliptical galaxy naturally suggests the
following {\it scenario}: 
\par (i) massive GCs on box orbits (in triaxial galaxies) or on low angular
momentum orbits (in axisymmetric galaxies) lose their orbital energy rather 
quickly;
\par (ii) after $\sim 500$ Myr many GCs, sufficiently robust to tidal
deformation, are limited to move in the inner galactic region where 
they merge and form an SSC;
\par (iii) stars of the SSC buzz around the nucleus where some of them are 
captured 
by a BH sitting there, partly increasing the BH mass;
\par (iv) part of the energy extracted from the SSC gravitational field goes 
into
e.m. radiation inducing a high nuclear luminosity up to AGN levels.
\par Point (i) has been carefully studied in \cite{PCV} and \cite{cap05}
in self consistent models of triaxial core-galaxies, and presently under study 
in triaxial 
cuspy-galaxies with dark matter halo \cite{caplecetal06}; the validity of point 
(ii) has 
been demonstrated by first results of \cite{cap93},
while the resistance to galactic tidal forces of sufficiently compact 
GCs confirmed by \cite{miocchi06} and the actual formation of an SSC via
orbitally decayed cluster merger has been proved by detailed N-body 
simulations \cite{capetal06}, \cite{mioetal06}.
Points (iii) and (iv) deserve a deeper investigation by mean of accurate
modeling, even if they seem reasonably well supported by previous studies 
\cite{cap93},\cite{cap04}.

\section{Conclusions}
\label{concl}
Various papers by our research group have shown that
many of the observed GCS features find a natural explanation in terms of
evolution of a GCS in the galactic field, assuming the (very conservative) 
hypothesis
it was initially radially distributed as the galactic stellar component and
coeval to it. In other words, no {\it ad hoc} assumptions are needed to explain, 
for
instance,  the difference, observed in many galaxies, among the GCS-halo star
profiles.
The initial presence of some massive GCs  ($M\geq 5\times 10^6$ M$_\odot$)
lead to the formation of a central SSC via merger of these orbitally
decayed massive clusters. The SSC mix it up with the galactic nucleus in which 
is
embedded and constituted a mass reservoire to fuel and accrete a massive object 
therein.
Observationally, this latter picture is supported by the observed positive
correlation between the estimated  quantity of mass lost by a GCS in galaxies
and the mass of their central BHs (see Fig. 1 in \cite{cap01}).
On the theoretical side, the modes of mass
accretion onto the BH via star capture from the merged SSC still remain to be
carefully investigated.

\section{Acknowledgments}
Thanks are due to P. Di Matteo, L. Leccese, P. Miocchi, A. Tesseri, and A. 
Vicari,
whose collaboration has been precious in a significant part of the work on the 
topics
discussed here.
%
%
%
%
%
%
%
\input{referenc}
\printindex
\end{document}

%% file: referenc.tex
%
%

%
%

%
%
%


